**Interfacial Hot Carrier Collection Controls Plasmonic Chemistry**

*Fatemeh Kiani[1], Alan R. Bowman[1], Milad Sabzehparvar[1], Can O.Karaman[1], Ravishankar Sundararaman[2], Giulia Tagliabue[1]\**

[1] *Laboratory of Nanoscience for Energy Technologies (LNET), STI, École Polytechnique Fédérale de Lausanne, 1015 Lausanne, Switzerland*

[2] *Department of Materials Science & Engineering, Rensselaer Polytechnic Institute, 110 8th Street, Troy, New York 12180, USA*

*\*E-mail: giulia.tagliabue@epfl.ch*

**Abstract**

Harnessing non-equilibrium hot carriers from plasmonic metal nanostructures constitutes a vibrant research field. It promises to enable control of activity and selectivity of photochemical reactions, especially for solar fuel generation. However, a comprehensive understanding of the interplay of plasmonic hot carrier-driven processes in metal/semiconducting heterostructures has remained elusive. In this work, we reveal the complex interdependence between plasmon excitation, hot carrier generation, transport and interfacial collection in plasmonic photocatalytic devices, uniquely determining the charge injection efficiencies at the solid/solid and solid/liquid interfaces. Interestingly, by measuring the internal quantum efficiency of ultrathin (14 to 33 nm) single-crystalline plasmonic gold (Au) nanoantenna arrays on titanium dioxide substrates, we find that the performance of the device is governed by hot hole collection at the metal/electrolyte interface. In particular, by combining a solid- and liquid-state experimental approach with *ab initio* simulations, we show a more efficient collection of high-energy d-band holes traveling in [111] orientation, resulting in a stronger oxidation reaction at the {111} surfaces of the nanoantenna. These results thus establish new guidelines for the design and optimization of plasmonic photocatalytic systems and optoelectronic devices.

**Introduction**

Prompt collection of photoexcited hot carriers in plasmonic metal nanostructures offer substantial promises for the development of applications such as tunable photodetection[1–3] and selective photocatalysis.[4–6] Practical realizations of hot carrier devices thus require a full understanding of plasmonic hot carrier-driven processes, including plasmon excitation (optical response), hot carrier generation, carrier transport and collection at solid/solid or solid/liquid interfaces. To date, despite significant experimental and theoretical investigations on plasmon excitation and hot carrier generation, electronic processes, i.e. transport and collection, have been less considered. Specifically, works have been able to unravel in detail the picture of the hot electron or hot hole collection at the solid-solid interface[1,2,7–9], but fewer studies have tried to resolve how the charge carrier collection occurs at the solid-liquid interface.[10–12] In fact, the majority of the focus has been on analyzing the external quantum efficiency (EQE) and the photo-induced activity of photocatalysts based on the plasmon resonance absorption[13–15] or on tracking molecular transformations via Raman spectroscopy.[16–18] Carriers generated by plasmon decay impinge upon the surface of a plasmonic nanostructure to be collected, either ballistically or after scattering against other carriers, phonons, or defects in the metal. These ultrafast (a few tens of femtoseconds to picoseconds) scattering processes thermalize the carriers and bring their energy distribution closer to the Fermi level of the metal.[19] However, plasmonic hot carrier applications require high energy electrons or holes to efficiently drive ensuing processes.

Plasmon-driven photocatalytic devices typically employ metallic nanoantennas/semiconductor heterostructures due to their efficient hot carrier separation. As a result, hot carrier-driven processes typically involve the complex interplay of carrier collection at both metal/semiconductor and metal/electrolyte interfaces. To gain complete system understanding, the geometry of the nanostructures and their absorption properties need to be precisely controlled. In particular, polycrystalline metal structures or films that have been used in almost all the plasmonic metal/semiconductor devices are not ideal for fundamental studies. Instead, single-crystalline nanoparticles or metal films, e.g gold micro-flakes[20,21], can be leveraged to obtain high-quality plasmonic nanoantenna (arrays) with unique optical properties[22–24] and well-defined crystallographic surfaces, which exhibit distinct catalytic properties[25,26] as well as distinct hot carrier transport properties.[27] Additionally, experimental quantification of the internal quantum efficiency (IQE) of hot carrier collection at each interface is critical to clarify the role of interfaces and their interplay, and to further elucidate the potential opportunities and limitations of hot carrier collection in these devices. Moreover, an appropriate model for hot carrier injection across metal/electrolyte interfaces has remained elusive. Overall, lack of comprehensive experimental approach, a theoretical model, and a highly-controlled nanostructure system have prevented an in-depth investigation on the hot carrier transport and collection processes occurring within hot carrier-driven photocatalytic systems.

In this work, we implement a combination of solid-state photocurrent and liquid–state photoelectrochemical measurements to simultaneously study the transport and collection of hot carriers across the metal/semiconductor and metal/electrolyte interfaces. Uniquely, we leverage highly-controlled single-crystalline plasmonic gold (Au) nanoantenna arrays on a $TiO_2$ semiconducting substrate. We use wavelength-dependent scanning electrochemical microscopy (photo-SECM) to locally probe the hot hole-driven photo-oxidation of a redox molecule at the nanoantenna surface under different excitation wavelengths across the entire visible regime (1.4-2.8 eV) and for different thicknesses of the nanoantennas. This spectral range spans across both the intra- and interband regimes of Au, and allows disentanglement of plasmon absorption and interband excitation effects. In combination with solid-state measurements, photo-SECM analysis reveals that the IQE of the photoanode device, in which both hot electrons and hot holes are collected, is controlled and limited by the hot hole collection. Our injection probability model for hot hole collection, retrieved from the combination of experimentally determined IQEs of the photoanodes with *ab initio* predictions of hot hole generation and transport, reveals the maximum extraction probability of the d-band generated high-energy holes particularly from the top {111} facet. Additionally, we find that the thinnest (16 nm) nanoantennas are most efficient for collecting higher energy d-band holes and importantly, suitable for ballistic collection of hot carriers. Overall, our highly-controlled experimental approach allows quantitative comparison with optical modeling and theoretical calculations, enabling the deconvolution of separate optical and electronic contributions in hot carrier-driven photocatalytic systems. This information and methodology can thus play a critical role towards better design and optimization of plasmonic catalysts.

**Results and Discussion**

We fabricated plasmonic photocatalytic devices (photoanode) consisting of an array of Au nanodisks (Au NDs) with sizes of the order of tens of nanometers on top of a $TiO_2$/ITO film on a fused silica substrates (**Figure 1a**). Uniquely, our nanoantennas are made from single-crystalline Au microflakes (SC Au MFs)[20] (see Methods) to exclude the influence of grain boundaries and thus reduce ohmic losses. Importantly, their atomically-smooth and well-defined {111} crystallographic surfaces allows us to deconvolute the effect of different crystallographic facets from other effects in plasmonic photocatalysis. We use $TiO_2$ because of its large optical band gap (EG ~ 3.2 eV[28]), preventing visible-light absorption within the $TiO_2$ film, and we note it forms Schottky barrier ($\Phi_B$) across the Au/TiO$_2$ heterojunction, enabling hot-electron collection. The Au NDs are in contact with an electrolyte containing a reversible redox molecule, Fe(CN)$_6^{4-}$ (Ferrocyanide, the reduced form, *Red*) which enables hot hole collection through photoelectrochemical oxidation reaction (**Figure1.a**). The oxidaton of this molecule proceeds via one-electron-transfer, outer-sphere mechanism with a fast kinetic.[29] Moreover, the chosen molecule does not absorb visible-light. For complementary solid-state studies, we also work on a plasmonic photodetector device (Schottky photodiode) consisting of an array of SC Au stripes on the same TiO$_2$/ITO film (see Methods), providing an ideal experimental platform for studying the hot electron collection from the same metal with the same semiconductor (**Figure1.b**). The stripe geometry is dictated by the need of a direct electrical contact to the plasmonic nanoantennas but we note that the electron injection, controlled by the properties of the interface, is identical in both devices. The nanoantennas were fabricated from Au MFs with varying thicknesses of 14 to 33 nm to study the effect of nanoantenna thickness on hot carrier generation, transport, and collection processes within our device.

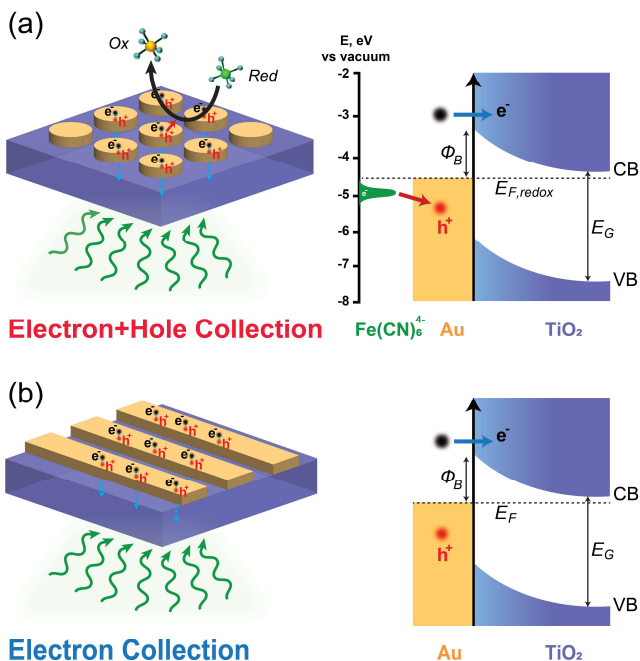

**Figure1.** Schematic of interfacial hot carrier collection in plasmonic metal/semiconductor heterostructure devices. (a) Au nanodisks (Au NDs)/TiO$_2$ photoanode in contact with a reductant (Red) molecule and band alignment showing hot electron and hot hole collection across the Au/TiO$_2$ and Au/Fe(CN)$_6^{4-}$ interfaces, respectively. (b) Au stripes /TiO$_2$ photodiode geometry and band alignment showing hot electron collection across the Au/TiO$_2$ interface.

Photo-SECM is a unique technique for ultra-sensitive[30] and fast detection[31] of tiny photo-chemical reactions on small-size nanostructures.[32–34] For photoelectrochemical experiments, we performed a series of photo-SECM measurements on Au NDs arrays with disk thicknesses of 16, 25, and 33 nm and average diameters of 68, 80, and 80 nm, respectively. We present the implemented photo-SECM approach in **Figure 2.a**. SEM image in **Figure 2.b** shows one of our fabricated NDs arrays for a 25nm-thick Au MF with a lateral size of 140 μm on a $TiO_2$/ITO-coated fused silica substrate. The SEM and AFM images in **Figure2.c** and **d** show the magnified view of the fabricated ND structure. SEM and AFM images of the other fabricated NDs array structures are shown in **Figure S3**. Thanks to the single crystallinity of the flakes, the fabricated nanoantennas exhibit exact shape, size, and ultra-smooth surfaces.

We determined absorption spectra of the structures experimentally and by simulation under front illumination in ambient air condition (see Methods and **Figure S4**). Measured absorptions are in good agreement with numerical simulations (**Figure S4**, dashed lines). To determine the absorption spectra with the illumination condition required for photoelectrochemical experiments, we implemented simulations for an aqueous medium and back illumination condition (see Methods). Calculated absorption spectra are plotted in **Figure 2.f**. All the Au NDs array heterostructures show a dipole plasmon resonance mode in the intraband region (1.57 to 1.75 eV) which enables us to disentangle plasmon absorption and interband excitation effects. Disentangling these two effects was challenging in previous studies as the resonance of the structure overlapped with the interband regime.[13–15]

For photoelectrochemical experiments, we used an aqueous electrolyte solution containing 4mM $Fe(CN)_6^{4-}$ and 0.25M KCl, as supporting electrolyte and positioned a 2.4 μm Pt UME tip at 2.5 μm away from the substrate. We performed photo-SECM measurements in competition mode, where an oxidation reaction proceeds at both the UME tip and the substrate (**Figure 2.a**). During the competition SECM experiment, we applied a potential of 0.4 V vs Ag/AgCl at the Pt UME tip corresponding to electro-oxidation of $Fe(CN)_6^{4-}$ at a diffusion-controlled rate (see **Figure S2.b**) while the substrate was illuminated from the bottom with a tunable, monochromatic light at different photon energies (see Methods). The substrate was at open circuit condition and grounded to take away the accumulated electrons in the ITO layer (**Figure 2.a**). We focused the laser beam on the Au NDs with a spot diameter of 30 μm to drive the photo-oxidation reaction only at the illuminated area of the substrate upon hot carrier generation and hot hole transfer at the Au/electrolyte interface. We measured the current through the tip as a function of excitation power to monitor changes in the local concentration of $Fe(CN)_6^{4-}$ and thus the photoelectrochemical dynamics at the plasmonic substrate. **Figure 2.e** shows the time-trace of the tip current ($i_{TiP}$) upon illumination of a 25nm thick ND array at an excitation wavelength of 450±10 nm, where intensity was modulated up to ~ 14 W/cm$^2$. To show the repeatability of the results, we performed the measurements three times at each illumination intensity. We note that once the light is turned off, $i_{Tip}$ returns to the baseline current ($i_{Tip,dark}$), indicating the rapid diffusion and charge transfer of the redox molecules within the tip-substrate gap that enables achieving a steady-state response easily. We observe a linear decrease in the magnitude of the $i_{Tip}$ relative to the dark current ($i_{Tip}/i_{Tip,dark}$) by increasing the excitation intensity (**Figure S8.b**). This decreasing trend indicates that the local concentration of Fe(CN)$_6^{4-}$ at the tip-substrate gap decreases due to a hole-driven oxidation reaction at the plasmonic substrate which gets enhanced in kinetics by illumination intensity. To confirm that the observed enhanced photo-oxidation at the Au NDs originates from the hot carrier generation and collection rather than photothermal heating effects in the range of the studied excitation intensities, we implemented a SECM approach[35,36] for probing the photothermal heating at a plasmonic substrate. Local heating at a plasmonic substrate results in enhanced mass transfer rates of the redox molecules, as well as a shift in the formal potential of a redox couple.[36] Our SECM measurements showed no sign of the temperature increase and thus local heating effects on the 25 nm-thick Au NDs array under illumination at its plasmonic resonance and excitation intensities up to 64 W/cm$^2$ (**Figure S6**).

To investigate the wavelength dependence of the photochemical oxidation rate of Fe(CN)$_6^{4-}$, we studied a broad excitation wavelength range of 450 to 832 nm covering the entire intra- and interband transition regimes for hot carrier generation in Au NDs. The measured $i_{Tip}/i_{Tip,dark}$ responses under different excitation intensities at each wavelength are shown in **Figure S8.b**. We implemented a diffusion model[13] to simulate calibration curves correlating $i_{Tip}/i_{Tip,dark}$, substrate photocurrent ($i_{sub,photo}$), and the reaction rate constant ($K_{eff}$) of the photo-oxidation reaction of Fe(CN)$_6^{4-}$ at the substrate. The substrate photocurrents were extracted using the calibration curve obtained by the diffusion model (**Figure S8.c**) and the measured data of $i_{Tip}/i_{Tip,dark}$ under different illumination intensities (**Figure S8.b**) for each excitation wavelength, and plotted as a function of power in **Figure S8.d**. As a control experiment, the same SECM measurement was performed on a bare TiO$_2$/ITO substrate in the absence of Au NDs. A slight enhanced photo-induced $i_{Tip}$ was observed at shorter wavelengths region by increasing the excitation intensity (**Figure S9**). To exclude the contribution from TiO$_2$ substrate, the TiO$_2$/ITO photocurrent was subtracted from the Au/TiO$_2$/ITO photocurrent (see Methods). Therefore, the substrate photocurrent plots in **Figure S8.d**, which are after the subtraction correspond solely to the hot hole-driven photo-oxidation reaction at Au NDs surfaces. The external quantum efficiency (EQE) of the photoelectrochemical reaction is determined from the slope of the linear fit to the $i_{sub,photo}$ vs illumination power curves for each excitation wavelength using **equation S14**. The linear relationship of $i_{sub,photo}$ vs power shows different slopes, clearly reflecting that the quantum efficiencies of the hot hole-driven oxidation reaction are wavelength-dependent. The measured absorption allows us to calculate the IQE, defined as the probability of a chemical reaction per absorption photon. The obtained EQE and IQE spectra are plotted in **Figure 2.g** and **h** for all the ND arrays (see **Figure S10** for normalized EQE and IQE spectra). We observe a common feature in all EQE spectra: a peak associated with the characteristic peak plasmon absorption energy in each structure. Conversely, the IQEs are not purely monotonic and interestingly all the three IQE curves exhibit an intermediate peak around 585 nm (2.12 eV) and the maximum efficiency was obtained in the interband region (> 2.4 eV, gray area in **Figure 2.h**), as observed in previous studies.[10–12] Comparing the thickness-dependent IQE spectra, we find a significant difference between the 16 nm and thicker ND structures, in particular in the interband region. The IQE of 16 nm ND structure increases monotonically from 2.4 to 2.8 eV, while the IQE of 25 and 33 nm ND structures rises up to a peak at 2.6 eV, and then drops to higher photon energies up to 2.8 eV. This is possibly because of an increased charge recombination at the Au/$TiO_2$ interface due to the concurrent charge generation in $TiO_2$ at higher photon energies, as also observed in previous studies.[37,38] Nonetheless, this interfacial recombination dose not impact the performance of the 16 nm-thick NDs. Thus, these observations suggest that decreasing the nanoantenna thickness would alter the hot carrier-driven processes across the interfaces. As both hot electron and hot holes are collected at different surfaces it is not clear how the efficiencies of these interfacial collection processes are affected by decreasing the thickness. Therefore, to clarify these two interface contributions, we first focus on the solid-state part to understand the role of metal/semiconductor interface.

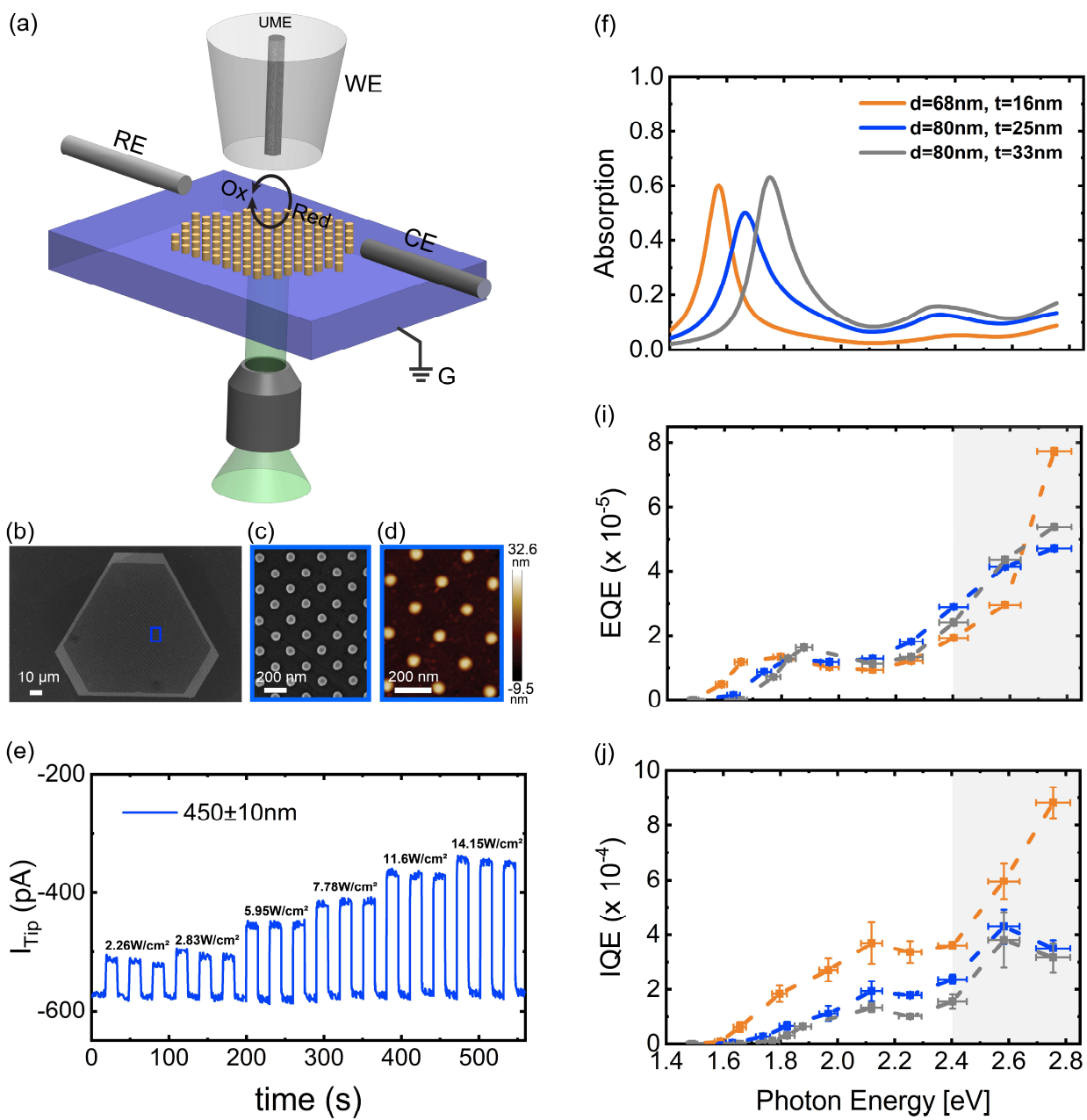

**Figure 2.** Liquid-state photoelectrochemical measurement results. (a) Schematic of the designed plasmonic heterostructure and photo-SECM configuration in a competition experiment mode. A fabricated Au NDs array from a SC Au MF on TiO$_2$/ITO substrate is in contact with an electrolyte contacting 4mM Fe(CN)$_6^{4-}$ (Red) and 0.25 M KCl. A 1.2 μm-radius Pt ultra-micro electrode (UME) tip is positioned 2.5 μm away from the substrate. The UME tip is biased at 0.4 V vs Ag/AgCl (reference electrode, RE), and the substrate is at open circuit and grounded. Light is incident on the plasmonic Au NDs array from the bottom. The same oxidation reaction happens at the tip electrode and substrate surface. The current is measured through the tip electrode (working electrode, WE). A Pt wire is used as a counter (CE) electrode

to complete the circuit. (b) SEM image of an entirely patterned 25 nm-thick Au MF. (c) Higher magnification SEM and (d) AFM images of the fabricated Au NDs array from the Au MF in (b). The average ND diameter and thickness are 80 and 25 nm, respectively. The array periodicity is 200 nm. (e) Time trace of tip current ($i_{Tip}$) obtained from the 1.2 µm-radius Pt UME upon illumination of the Au NDs array in (b) with the excitation wavelength of 450±10 nm at different light intensities. (f) Simulated absorption spectra of the fabricated heterostructures having different Au ND thicknesses of 16, 25, and 33 nm exhibiting resonance peaks at 1.57, 1.66, and 1.75 eV, respectively. (i) Experimentally determined EQE and (j) IQE spectra for the same heterostructures. The gray shaded areas depict the purely interband region and the dashed lines are a guide to the eye in panels (i) and (j).

To study hot electron collection in solid-state, we fabricated a plasmonic Schottky photodiode device consisting of an array of Au stripes with a thickness of 14 nm, close to the thickness of the best performing ND device on the same $TiO_2$/ITO film. We present the implemented solid-state approach in **Figure 3.a**. SEM image in **Figure 3.b** shows a fabricated 30x30 µm$^2$ stripe array. The magnified SEM and AFM images (**Figure 3.c** and **d**) show the fabricated stripe structure with the width of 70 nm and periodicity of 230 nm. From **Figure 3.b** we observe the stripe array is connected to the Au flake body, which can enable photocurrent measurements by electrically contacting microprobes on a non-patterned flake area and on a sputtered Ohmic contact to $TiO_2$. The photocurrent can be collected while illuminating the sample through the bottom with a tunable, monochromatic light (see the schematic in **Figure. 3a** and Methods). **Figure 3.e** shows the time-trace of the short circuit current ($I_{SC}$) upon illumination of the device with wavelengths from 450 to 840 nm at different powers. From the *I-V* response shown in **Figure 2.e**, a highly rectifying behavior was observed. Fitting of this plot to the diode equation[40] allowed the estimation of a Schottky barrier of ~ 1.25 eV across the Au/$TiO_2$ heterojunction. We determined the EQE and absorption spectra of the device experimentally by measuring both the wavelength-dependent photocurrent as well as transmission and reflection spectra under the same light polarization condition (see Methods) and plotted them in **Figure 3.f**. A resonance peak at 665 nm (1.85 eV) appears in both spectra. This feature disappears when light polarization is parallel to the stripes (see **Figure S5**). Such behavior confirms that the photocurrent originates from optical excitation of the dipolar plasmon mode in the nanoantennas.[1] The observed similar EQE trend with the absorption spectrum indicates the plasmon excitation manifests itself in an enhanced EQE of the device. Interestingly, consistent with the previous solid-state works[1,2], we observed that the IQE spectrum (**Figure 3.j**) exhibits a spectral feature peaking at 550 nm (2.25 eV). The asymmetry between energy distributions for hot carriers generated by interband transitions[19] (peak of distribution at lower energies close to the Fermi level ($E_F$) for hot electrons), combined with our Schottky barrier height (1.25 eV) has consequences for hot electron collection efficiency across Au/TiO$_2$ interface. As a result, we observe an abrupt drop in the solid-state IQE at energies above the interband threshold (gray area in **Figure 3.j**). On the other hand, due to the generation of high-energy d-band holes, there is still possibility of hole collection at Au/electrolyte interface, resulting in a continued IQE growth in the entire interband regime for the best performing device (c.f. **Figure 2.j**, orange dashed curve, 2.4 to 2.8 eV). Comparing the liquid-state photoelectrochemical and solid-state photocurrent measurement results, we find a different magnitude and trend between the experimentally obtained IQEs for the two devices. Higher magnitude of the IQE for the hot electron photodetector device and a very different IQE trend in particular in the interband regime imply that the IQEs of the photoanodes are not limited by the hot electron collection, rather by hot hole collection.

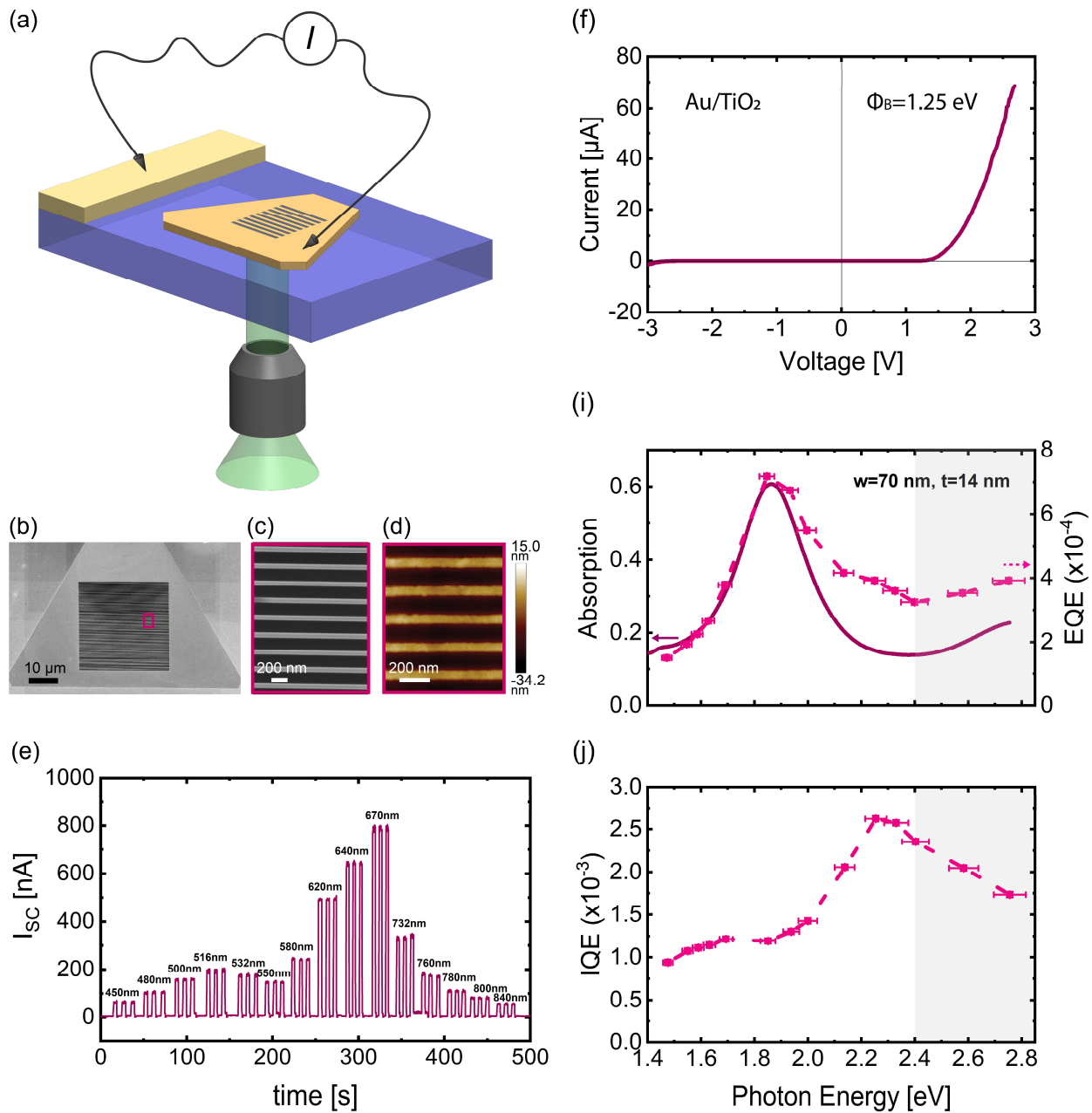

**Figure 3.** Solid-state photocurrent measurement results. (a) Schematic of the designed plasmonic heterostructure and measurement configuration. A stripe Au pattern is fabricated from a SC Au MF on TiO$_2$/ITO substrate together with a 100 nm-thick sputtered Au film Ohmic contact. Light is incident on the plasmonic Au stripe array from the bottom and the photocurrent is collected by two microcontact probes electrically connected to the Au flake and the sputtered Au contact pad. (b) SEM image of a 30x30 µm$^2$ stripe array from a 14 nm-thick Au MF. (c) Higher magnification SEM and (d) AFM images of the fabricated Au stripe array from the Au MF in (b). The average stripe width and thickness are 70 and 14 nm, respectively. The array periodicity is 230 nm. (e) Time-trace of the short-circuit current ($I_{SC}$) for the
11

fabricated stripe array in (b) upon illumination with the excitation wavelengths of 450±10 to 840±10 nm. (f) Measured I-V plot of the fabricated heterostructure showing a metal-semiconductor Shottky diode behavior across the Au/TiO$_2$ interface. A Shottky barrier height of $\Phi_B$=1.25 eV was estimated after fitting these data. (i) EQE (purple dashed curve) and absorption spectra (purple solid curve) of the fabricated heterostructure exhibiting the EQE peak response at the plasmon resonance of the structure at 1.85 eV. (i) Experimentally determined IQE spectrum of the same heterostructure. The gray shaded areas depict the purely interband region and the dashed lines are a guide to the eye in panels (i) and (j).

To be able to understand how the hot hole collection limits the IQE in the photocatalytic system, we leverage theory to disentangle the mechanisms controlling the generation, transport, and injection of the hot carriers across the metal/electrolyte and metal/semiconductor interfaces. **Figure 4.a** shows the calculated spatially-resolved absorption 2D profiles, i.e. hot carrier generation profiles, in Au NDs with thicknesses of 16 and 33 nm on TiO$_2$/ITO/glass substrate at their plasmon resonance (intraband regime) and at photon energy of 2.75 eV (450 nm, interband regime). We observe a uniform hot carrier generation across the whole volume of the 16 nm ND close to the both solid-solid and solid-liquid interfaces while a localized generation at the solid-solid interface of the 33 nm ND both for on-resonance and off-resonance excitations. We then employed *ab initio* simulations[41] to elucidate the impact of carrier transport and separate the contributions from carriers based on the number of times they are scattered before collection (*N*) (see Methods). The energy-resolved flux, $F_N(E)$, of hot carriers with energy above (hot electrons) and below (hot holes) the $E_F$ that can reach all interfaces was calculated at each photon energy. As the fabricated structures are single crystalline, no orientational averaging was applied on the ballistic hot carrier fluxes. We calculated the carrier fluxes for {111} top and bottom and {110} side facets[20] separately. **Figure 4.b** compares the spatially-resolved carrier fluxes of hot holes with energy greater than a threshold (-0.15eV, close to the highest occupied molecular orbital (*HOMO*) level of Fe(CN)$_6^{4-}$ (-0.19 eV) or -2eV, to highlight the presence of the high-energy hot holes) for 16 and 33-nm thick Au NDs at 450 nm and their plasmon resonance wavelength. Under on-resonance illumination of the structures in the intraband regime, the maximum flux of the holes that can reach the interface for the 16 and 33 nm NDs are comparable, however, we see a clear difference in their spatial distribution. For the 16 nm ND, the flux is well distributed across the interfaces, while a localized flux is observed on the side surface close to the bottom interface of the thicker ND, as is evident from their corresponding calculated energy-resolved carrier fluxes at different surfaces shown in **Figure S11**. At energies above the interband threshold, accessible by the 450 nm illumination, the carrier flux become more uniform also for the 33 nm ND due to the improved generation profile (**Figure 4.a** at 450 nm). But still, 33 nm ND exhibits a much smaller maximum flux for both energy thresholds compared to the 16 nm ND, in particular for high-energy holes due to the losses in the transport. Therefore, due to the observed non-uniform carrier generation and flux distribution, as well as the increased distance that holes must travel before reaching the Au/electrolyte interface in the thicker nanoantennas, a slower IQE growth was observed compared to the 16 nm ND (**Figure 2.j**). This is more critical for collecting the high-energy d-band holes, where the mean free-path of hot holes decreases to just a few nanometers[18,40], leading to a nonmonotonic behavior of thicker ND IQEs upon transition from partially to entirely interband excitation (**Figure 2.j**, 2.1 to 2.8 eV). **Figure 4.c** maps the mean free-path of holes in the top d-band in Au across velocity directions. The unit vector of velocity directions of these holes are distributed non-uniformly across the unit sphere with much higher probability between the [110] and [111] directions. These most probable velocity directions also have longer mean free-paths, indicating that the anisotropy favors hot hole transport towards the top and sides of our single-crystalline samples (compared to intermediate directions).

Finally, we combined our experimental IQE results with our generation and transport model to extract the injection probability across the Au/electrolyte interface. Specifically, by assuming the injection probability, $P_{inj}(E)$, is constant for all samples (as the same surfaces are exposed), we can state that for the NDs:

$$IQE = \int_{-\infty}^{HOMO} F_N(E) \cdot P_{inj}(E) dE \quad (1)$$

Where $E$ is the energy of the hot holes with respect to $E_F$ and $N$ is the number of the scattering events we include in modeling. As we know IQE and $F_N(E)$, we developed a stochastic fitting model to extract $P_{inj}(E)$. Specifically, we carried out a fitting procedure several hundred times to calculate the average and possible spread of $P_{inj}(E)$ (see Method for further details). The resulted $P_{inj}(E)$ from this fitting approach for the top and side surfaces is shown in **Figure 4.d**. No significant difference was obtained in injection probability between non-scattered (**Figure 4.c**, N=0) and scattered (**Figure S12**, N=4) holes, as the flux distribution of initial and homogenized carriers is the same at energies below -2 eV (**Figure S13**). We observe the extraction probability approaches an exponentially growing curve for increasing hole energy, with around 5 times higher probability of -2.9 eV holes collected from the top compared to the ones collected from the side surfaces. A very low upper-bound injection probability of ~ 0.1 % is obtained for intraband generated holes with energy of -1.46 eV, although they could have sufficient energy to participate in *HOMO* electron transfer due to the very low oxidation energy of Fe(CN)$_6^{4-}$ (-0.19 eV, **Figure S2**). We propose that higher energy holes are more strongly oxidizing and more likely to participate in this reaction due to a tunneling barrier that they must overcome in order to reach the physisorbed molecule on the surface.[42] This indicates that mostly the d-band generated high-energy holes strongly participate in the oxidation reaction in our system, in agreement with suggestions from previous studies on hot hole-driven reactions.[10–12] Therefore, it is expected that these higher energy holes are mostly collected ballistically, otherwise they immediately lose their energy under scattering events and would have insufficient energy to drive the reaction. **Figure 4.e** shows the predicted IQE spectra from $F_0(E)$ for the holes that reach to the top and side surfaces and the obtained $P_{inj}(E)$ in **Figure 4.d** together with the experimentally determined IQEs for 16 and 25 nm-thick Au NDs heterostructures (see 33 nm-thick NDs in **Figure S14**). Our transport model and fitting approach for injection probability reproduced our experimental data well, in particular our best performing device, 16 nm ND heterostucture. Almost the same computed IQE trend was obtained for the scattered hot holes (**Figure S14**).

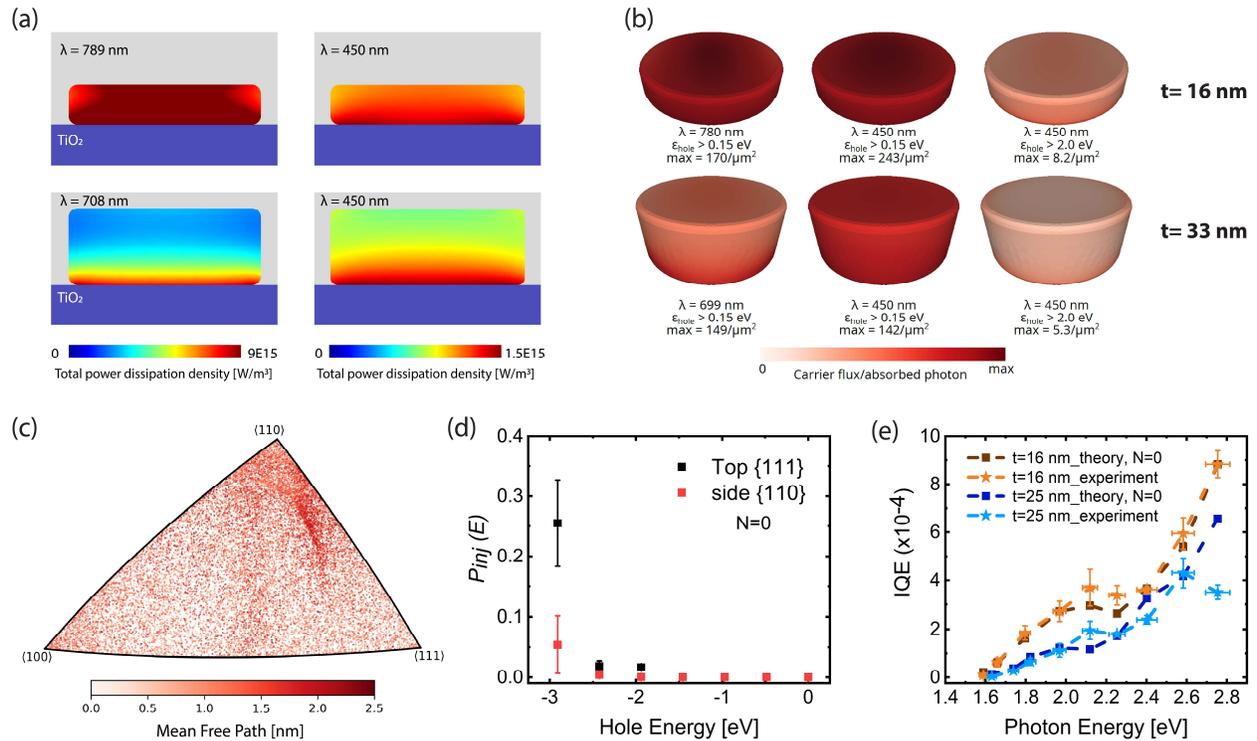

**Figure 4.** Hot carrier generation, transport and injection in plasmonic heterostructure devices. Calculated spatially-resolved (a) 2D absorption profiles and (b) hot hole fluxes reaching the interfaces greater than 0.15 and 2 eV under illumination at 450 nm and the resonance wavelength of the Au NDs having thicknesses of 16 and 33 nm. Fluxes are normalized to the absorbed photon and unit area of the surface.

(c) Distribution of velocity unit vectors of holes in the top d-band of Au, colored by mean free path, and shown within the irreducible wedge of the unit sphere under cubic symmetry (bounded by the [100], [110] and [111] crystallographic orientations as vertices). (d) Injection probability ($P_{inj}$) for hot holes collected from the top {111} and side {110} facets. The $P_{inj}$ plots are extracted from the fitting approach using the energy-resolved hot hole fluxes and experimentally determined IQEs of Au NDs heterostructures. (e) Calculated IQE spectra based on energy-resolved hot hole fluxes and estimated $P_{inj}$ in (d) for non-scattred carriers (N=0) together with the experimentally determined IQEs for 16 and 33 nm-thick Au ND heterostructures. Dashed lines are a guide to the eye.

Based on our analysis of the injection probability, we also calculate the maximum possible IQE for our system that we could achieve for two upper-bound assumptions. First, we assume that the illumination and chemical reaction happen on the same side (e.g. bottom surface). The maximum IQE, in this case, is 0.12 % (**Figure S15.a**), which is still close to the experimentally obtained maximum value of 0.094 % in our system (**Figure 4.e**). Second, we assume all the generated holes immediately reach the top surface and are collected before any scattering events without any transport losses. The maximum IQE, in this case, is 5 % (**Figure S15.b**), which is ~ 50 times higher than our obtained value. This emphasizes that transport is one of the main limiting factors in the performance of hot carrier devices which should be taken into account in future designs.

**Conclusion**

In summary, we established a fully controlled system to experimentally quantify the IQE in ultrathin (14 to 33 nm) single-crystalline plasmonic Au nanoantenna array Schottky photodiode and photoanode devices that operate via the collection of hot electrons (Au/TiO$_2$) and both hot electrons and holes (Fe(CN)$_6^{4-}$/Au/TiO$_2$), respectively. All the Au nanoantenna array heterostructures were designed and fabricated to have plasmon resonance in the intraband region, allowing us to uniquely disentangle plasmon absorption and interband excitation effects. Our experimental IQE data combined with ab-initio modeling revealed the role of intra- and interband decay processes and carrier transport over the nanometer size of the antennas, proposing an injection probability estimation for hot holes collected at the metal/electrolyte interface. Comparing the measured IQE spectra for the two devices having very close thicknesses (14nm and 16 nm) showed a different magnitude and trend in IQEs, with an abrupt drop for the photodiode device and a continuous increase for the photoanode device in the interband regime. The magnitude difference suggested that the efficiency of the photoanodes is indeed controlled and limited by the hot hole collection at the metal/electrolyte interface. Our injection probability model indicated that mostly the d-band generated high-energy holes participated in the oxidation reaction in our photoanode system and particularly are collected from the top {111} surface. Our transport model and fitting approach for injection probability showed that these hot holes and electrons are mostly collected ballistically at both the $Fe(CN)_6^{4-}$/Au and Au/$TiO_2$ interfaces in our device with the highest collection efficiency for the thinnest nanoantenna photoanode device (16 nm) as compared to the thicker counterparts (25 and 33 nm). Our results and combined approach could reveal mechanistic insights into the generation, transport, and injection of hot carriers in hot carrier-driven photocatalytic systems, and be a guideline for the design of efficient devices operating in ballistic regime, e.g. plasmon-driven artificial photosynthetic systems or optoelectronics.

**Methods**

**Synthesis and device fabrication.** In order to make a transparent semiconducting substrate, first a 50 nm-thick ITO film was deposited on a cleaned 1.5x1.5 $cm^2$, 520 µm-thick fused silica substrate in a sputtering system (Pfeiffer SPIDER 600). Subsequently, a 40 nm $TiO_2$ film was deposited onto the ITO surface with Alliance-Concept DP 650 sputtering instrument. The deposited $TiO_2$/ITO film was then annealed in air at 450 °C for 2 hours. Large area SC Au MFs were directly grown on borosilicate glass substrates by a halide and gap-assisted polyol process.[20] A PMMA wet-transferring method[43] was used for transferring the Au MFs onto the $TiO_2$/ITO-coated glass substrate. The sample was then exposed to an oxygen plasma (4 minutes, 500 W) to remove any PMMA residue left from the transferring step. Next, a layer of PMMA 495k A4 was spin-coated on the sample (120 nm) and backed for 5 minutes at 180 °C. Then, to fabricate the photoanode device, electron-beam lithography (EBPG5000ES system) was used to write the nanodisk array pattern on the Au MFs (100 pA beam current with exposure of 425 µC/$cm^2$) following by PMMA development (sequentially immersing in MIBK and IPA solutions for 1 min each). It is to mention that due to the transparent sample, two 5nm/100nm-thick Ti-Pt stripes were deposited at two sides of the substrate as the reflective layer for providing a correct height measurement to have the electron beam focused on the substrate surface during the writing. Next, ion beam etching (Veeco Nexus IBE350) was used to etch the Au flake area around the nanodisks (ultra low IBE process, -10° angle). The Remained PMMA on top of the nanodisks was removed by immersing the sample in pure acetone (20 min) and isopropanol (3 min) followed by rinsing with DI water and drying with $N_2$. Lastly, the sample was exposed to a mild oxygen plasma (30 s, 350 W) for further cleaning the chemical residues. To fabricate the photodiode device, focused ion beam (FIB) milling (30 kV $Ga^+$ beam) was used to make an 30x30 µ$m^2$ Au stripe array from a transferred SC Au MF onto a TiO$_2$/ITO coated glass substrate. The FIB milling was performed with a FIB/SEM dual-beam instrument (Zeiss CrossBeam 540). A narrow film of Ti/Au (5 nm/100 nm) Ohmic contact was then deposited on the substrate surface close to the fabricated Au stripe array structure with a sputtering system (Alliance-Concept DP 650).

**Optical measurements.** To record reflectance and transmittance spectra of the structures, an inverted microscope (Nikon Eclipse Ti2) was used in combination with a grating spectrometer (Princeton Instruments Spectra Pro HRS-500) equipped with a Peltier-cooled 2D CCD detector (Princeton Instruments PIXIS 256). A fiber-coupled broadband laser-driven white light source (Energetiq LDLS™) focused on the back focal plane (BFP) of a long working distance, high-NA objective (Nikon 60x, NA=0.7) to illuminate the sample from the bottom with a collimated light. The reflected light was directed to the spectrometer. The measured reflectance was normalized to the reflectance of a silver mirror (ThorLabs, PF10-03-P01) and multiplied by the known reflectance of the mirror. Also, the background was subtracted from all the measurements. The transmittance measurements were carried out by top illumination of the sample through a bright field condenser lens. The sample was faced down for the transmission and faced up for the reflection measurements, ensuring the beam first hit the substrate and then Au nanostructures to implement the practical illumination condition required for the liquid-state and solid-state experiments. The reverse direction was applied for the front illumination condition.

**Photoelectrochemical and photocurrent measurements.** To perform liquid-state photoelectrochemical measurements, a custom-built photo-SECM set-up is integrated by adding to an inverted optical microscope (Nikon Eclipse Ti2), a home-built electrochemical reaction cell, a bi-potentiostat (Biologic SP-300), and an ultramicroelectrode (UME) tip connected to a micro/piezo scanner assembly (MMP1/Nano-F450, Mad City Labs ). Pt UME tips were fabricated by heat-sealing and hard pulling of a Pt micro-wire (25 µm, Goodfellow) within a borosilicate glass capillary (1mm ID, 0.5mm OD, Sutter Instruments) with a laser puller machine (P-200, Sutter Instruments), followed by physical contacting of a Cu wire to the Pt wire using a silver epoxy. The laser-pulled UMEs were perfected on an ultrafine polishing plate (BV-10, Sutter Instruments) under video monitoring. A Pt wire (0.5 mm diameter) and a leak-free Ag/AgCl electrode (LF-1, 1 mm OD, Innovation Instrument) were used as counter and reference electrodes, respectively. Potassium hexacyanoferrate (II) trihydrate (K$_4$Fe(CN)$_6$.3H$_2$O, 99.95%),  Potassium hexacyanoferrate (II) (K$_3$Fe(CN)$_6$, 99.98%) ,and potassium chloride (KCl, 99%) were purchased from Sigma-Aldrich and used as received. A high power supercontinuum white light laser (NKT Photonics) was used for plasmon excitation.

A tunable wavelength filter (SuperK VARIA) was employed to modulate the excitation wavelength and power in a wide spectral range of 450 to 840 nm with the bandwidth of 20 nm. An optical shutter (SH1, Thorlabs) was used to chop the incident light. A combination of two lenses was added to the optical path to provide a bottom illumination on the sample with a 30 µm diameter collimated beam. To define the beam diameter, first a CCD image of the laser spot was recorded and then fitted by a two-dimensional Gaussian. Spot diameter here is as the intensity that falls to $\frac{1}{e^2}$ of the maximum intensity. The solid-state photocurrent measurements were carried out by utilizing piezoelectric microprobes (Imina Technologies, miBots™) to electrically connect the sample and record the short-circuit current. Note that a polarizer (WP25M-VIS,Thorlabs) was added to the optical path to provide a polarized beam perpendicular to the Au stripes. The current-voltage (*I-V*) and time-trace of the photocurrent (*I-t*) curves were recorded through a Keithley 2450 SourceMeter.

**Numerical simulation.** The electromagnetic simulations were performed using the RF module of the COMSOL Multiphysics v5.6 to simulate absorption spectra as well as 3D internal electric field distributions across the volume of the nanoantennas. The latter was used as an input in the subsequent hot carrier transport calculation code. A 3D unit cell model, consisting of one Au nanoantenna (disk or stripe geometry) on $TiO_2$/ITO/fused silica substrate surrounded with a top layer of water (for disk) or air (for stripe), was simulated by setting the unit cell width equal to the array periodicity (200 nm for disks and 230 nm for stripes) and unit cell length equal to 200 nm and 300 nm for disks and stripes, respectively. Perfect magnetic conductor and perfect electric conductor boundary conditions were used at the sidewalls of the unit cell. A port boundary condition with the excitation "ON" was set at the bottom of the unit cell for the back illumination with a normal incident plane wave (450-850 nm) with electric field polarization perpendicular to the stripe length axis as well as for recording the reflected wave. A second port boundary condition without excitation was used at the top of the unit cell to record the transmitted wave. The absorbed power was calculated by volume integration of the electromagnetic power loss density over the nanoantenna volume. The wavelength-dependent complex refractive indices for SC Au, $TiO_2$ and ITO were taken from refs [44], [45], and [46]. A 2D diffusion COMSOL model was also implemented to simulate the tip current ($I_{Tip}$) response in photo-SECM experiments. Simulation details are provided in the **Supplementary information 4**.

**Hot carrier transport and injection predictions.**

Starting from the electric field distribution from the electromagnetic simulations, we predict the initial spatially-resolved energy distribution of hot carriers and their subsequent transport accounting for electron-electron and electron-phonon scattering using the Non-Equlibrium Scattering in Energy and Space (NESSE) simulation framework.[41] Briefly, this technique uses *ab initio* simulations of the optical excitation due to direct and phonon-assisted transitions to predict the initial carrier distribution, and then evolves spatially-resolved carrier energy distribution using the Boltzmann transport equation with a collision integral parameterized to first-principles electron-phonon and electron-electron scattering. The NESSE formulation predicts the carrier flux incident on the surfaces of the structure before scattering, after scattering once, twice, etc., allowing us to separate the contributions due to scattered and unscattered carriers. See Ref. 41 for a detailed description of the algorithm and the underlying first principles calculations used to parameterize the carrier generation and transport parameters.

In the main text the IQE is measured for a range of situations and the fluxes at the surfaces are calculated as $F_i(E)$, where subscript i corresponds to situation i (denoting excitation wavelength and nanostructure excited). We assume that the probability of a hole transferring from the nanostructure and causing a chemical reaction, $P_{inj}(E)$, is only a function of the hole's energy and not of the excitation wavelength or nanostructure shape (noting we assume transfer occurs at the exposed top {111} and side {110} surfaces). In this case we can state that

$$IQE_1 = \int F_{top_1}(E)P_{inj,top}(E) + F_{side}(E)P_{inj,side}(E)dE$$
$$IQE_2 = \int F_{top_2}(E)P_{inj,top}(E) + F_{sid}(E)P_{inj,side}(E)dE$$
$$\vdots$$

for all i situations, where the integral is over all hole energies and top and side refer to the fluxes and injection probabilities at the top and side.

We adopted a stochastic fitting approach to find the form of $P_{inj}(E)$ which best reproduced the IQEs. Specifically, we split the hole energy space into M distinct, equally spaced points (e.g. $E_{minimum}, E_{mean}, E_{maximum}$ for M=3). At each of these points we randomly assigned a value to $P_{inj}$ between 0 and 1, and between these points we carried out a linear interpolation to find $P_{inj}$ at all energies. We always treated top and side injection probabilities separately. We calculated the difference between computed and measured IQEs. We then randomly selected one of the defined points in energy space and varied this point until the best possible agreement between experiment and theory was found (within a tolerance). We repeated this approach, varying different defined points, until the difference between computed and measured IQEs was no longer changing. This gave us a set of $P_{inj}$ values which reproduced experimental data well. Finally, we repeated this fitting process 200 times, to obtain a spread in the $P_{inj}$ values which reproduce the experimental data well. In the main text we present the average and standard deviation of $P_{inj}$ values.

To carry out this fitting approach it is necessary to select a value for M, the number of distinct points to fit. Here we selected some M, carried out the above algorithm, and then calculating the average difference between calculated and measured IQEs. We then increased M until the difference between calculated and measured IQEs was not significantly changing (< 5 %) when M was further increased. In our simulations we found M=7 reproduced the experimental data well.

**Supporting Information**

Additional information about characterizatio of the redox molecule/Au/TiO$_2$ system; characterization of the Au nanoantenna heterostructures; SECM approach for probing the photothermal heating at Au NDs heterostructures; modeling the tip current response in photo-SECM; detailed analysis of IQE, transport and injection probability results.

**Acknlowledgements**

The authors acknowledge the support of the Swiss National Science Foundation (Eccellenza Grant #194181). The authors also acknowledge the support of the following experimental facilities at EPFL: Center of MicroNanoTechnology (CMi), and Interdisciplinary Centre for Electron Microscopy (CIME). Calculations were performed at the Center for Computational Innovations at Rensselaer Polytechnic Institute.

**Author Information**

Corresponding Author

Giulia Tagliabue, Ecole Polytechnique Federale de Lausanne (EPFL), Lausanne, Switzerland; orcid.org/0000-0003-4587-728X; Email: giulia.tagliabue@epfl.ch

Other Authors

Fatemeh Kiani, Ecole Polytechnique Federale de Lausanne (EPFL) , Lausanne , Switzerland; orcid.org/0000-0002-2707-5251

Alan R. Bowman, Ecole Polytechnique Federale de Lausanne (EPFL) , Lausanne , Switzerland; orcid.org/0000-0002-1726-3064

Milad Sabzehparvar, Ecole Polytechnique Federale de Lausanne (EPFL) , Lausanne , Switzerland, orcid.org/0000-0001-5594-6889

Can O. Karaman, Ecole Polytechnique Federale de Lausanne (EPFL) , Lausanne , Switzerland.

Ravishankar Sundararaman, Rensselaer Polytechnic Institute, New York, USA, orcid.org/0000-0002-0625-4592

**Table of Contents Graphic**

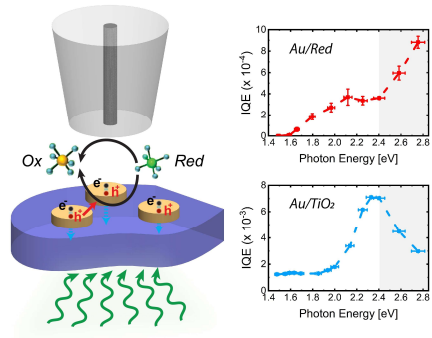